\documentstyle[osa,aplop,preprint,epsfig]{revtex}                               

\begin{document}
\preprint{UDEM-GPP-TH-96-35}
\vskip.5truecm
\title{Solitons in a Baby-Skyrme model with invariance
under area preserving diffeomorphisms}

\author{T. Gisiger and M. B. Paranjape}

\address{Groupe de physique des particules, Laboratoire de physique nucl\'eaire, 
Universit\'e de Montr\'eal, C.P. 6128, succ. centre-ville, Montr\'eal, Qu\'ebec,
Canada, H3C 3J7}

\maketitle

\begin{abstract}
We study the properties of soliton solutions in an analog of the Skyrme model 
in
2+1 dimensions whose Lagrangian contains the Skyrme term and the mass term, but
no usual kinetic term. The model admits a symmetry under area preserving
diffeomorphisms. We solve the dynamical equations of motion analytically for 
the case of spinning isolated baryon type solitons. We take fully into account
the induced deformation of the spinning Skyrmions and the consequent
modification of its moment of inertia to give an analytical example of related 
numerical
behaviour found by Piette {\it et al.}~\cite{1}. 
We solve the equations of
motion also for the case of an infinite, open string, and a closed annular
string. In each case, the solitons are of finite extent, so called ``compactons",
being exactly the
vacuum outside a compact region. We end with indications on the scattering of
baby-Skyrmions, as well as some considerations as the properties of solitons on
a curved space.
\end{abstract}

\section{ Introduction}

The baby-Skyrme model is a useful laboratory for studying soliton
physics\cite{1,2,3}.
It is the 2+1 dimensional analog of the model which describes the low energy
chiral dynamics of QCD\cite{5}, the usual Skyrme model\cite{4}. The model has 
direct applications in condensed
matter physics\cite{6} where baby-Skyrmions give an effective description in 
quantum Hall systems. In such systems, the dynamics are governed by the spin 
stiffness term, the Coulomb interaction and the Zeeman interaction. In the 
relativistic analog of this system the baby Skyrme model is governed by the 
Lagrangian\cite{1}
\begin{equation}
L = {1\over 2}\int d^2 x\, \Biggl[f_{\pi} \; \partial_\mu\vec\phi\cdot
\partial^\mu\vec\phi- {1\over 2}
\biggl(\partial_\mu\vec\phi\times\partial_\nu\vec\phi\biggr)^2 -
\mu^2 (\hat n-\vec\phi)^2\Biggr]\label{LBS}
\end{equation}
where $\vec\phi$ is a unit scalar iso-vector field, $\vec\phi\cdot\vec\phi=1$. 
This theory admits stable topological soliton solutions.
The kinetic energy corresponds to the spin stiffnesss term and the mass term 
corresponds to the Zeeman interaction, the correspondence being exact for the 
static sector. The Skyrme term (with four derivatives) is analogous to the 
Coulomb term (also with four derivatives) and both serve to stabalize from 
collapse topological configurations which yield Skyrmion solutions.   
In this paper we will consider the model arising in the limiting case when
$f_{\pi}\rightarrow 0$. This corresponds equivalently to $\mu\rightarrow +\infty$ with an appropriate rescaling. In the quantum Hall system, this corresponds to the limit $B\rightarrow +\infty$ (at finite Land\'e $g$ factor):
\begin{equation}
L = -{1\over 2} \int d^2 x \Biggl[{1\over 2}
\biggl(\partial_\mu\vec\phi\times\partial_\nu\vec\phi\biggr)^2+
\mu^2 (\hat n-\vec\phi)^2\Biggr]\label{LBSST}.
\end{equation}
We have already introduced this model in our previous letter\cite{7}, it is 
our aim in this paper to further elaborate on its properties. 

This type of model has been studied by Tchrakian {\it et al}~~\cite{8} for a 
complex
scalar field, with a somewhat modified potential. Here, for the static sector
they have essentially proven that the model is integrable. In their model there
exists a Bogomolnyi bound which is saturated and renders the differential
equation to be solved first order. As we shall see below there are certain
similarities between the model treated here and their model, however, the
corresponding Bogomolnyi bound here is not saturated. Hence it is not evident
that our model is also integrable for the static sector.

This Lagrangian admits the following symmetries. It is clearly Poincar\'e
invariant (Lorentz and translational invariance). In addition, because the
mass term picks the vacuum direction $\hat n$, the $O(3)$ invariance is
explicitly broken to $O(2)$, which corresponds to isorotations about $\hat n$:
\begin{equation}
\vec\phi\rightarrow R_3\cdot\vec\phi\qquad\qquad R_3\cdot\hat n = \hat n
\end{equation}

A more interesting symmetry comes from invariance under area
preserving diffeomorphisms for the static energy
\begin{equation}
E = {1\over 2}\int d^2 x \Biggl [ {1\over 4} \Bigl( \epsilon_{ij} 
\partial_i\vec\phi\times
\partial_j\vec\phi \Bigr)^2 + \mu^2 (\hat n-\vec\phi)^2\Biggr ]
\end{equation}
where we see that the vacuum configuration $\vec\phi=\hat n$ has zero energy.
Now if we change the coordinates $x^i\rightarrow {x'}^i$ with an area preserving
diffeomorphism
\begin{equation}
det \Biggr( {\partial {x'}^i\over\partial x^j}\Biggr) = 1
\end{equation}
then the integration measure does not change and 
\begin{equation}
\epsilon_{ij} \partial_i\vec\phi\times\partial_j\vec\phi \rightarrow det\Biggr(
{\partial {x'}\over\partial x}\Biggr) \epsilon_{ij}
\partial_i\vec\phi\times\partial_j\vec\phi 
\end{equation}
which also does not change since the determinant is 1. This symmetry is 
infinite
dimensional, therefore there is an infinite dimensional degeneracy in the
energies of the solutions. Intuitively each solution can be deformed like an
incompressible fluid to any shape imaginable. This should also translate into 
an infinite degeneracy of the ground state in the quantum theory, which is a 
persistent feature of microscopic models describing quantum Hall systems. 

We end this section by writing the static energy as follows:
\begin{equation}
E = {1\over 2}\int d^2 x \Biggl[ \Biggl( {1\over 2}\;\epsilon_{ij}\,
\partial_i\vec\phi \times \partial_j\vec\phi \pm \mu (\hat n-\vec\phi) 
\Biggr)^2
\pm 2 \mu B(\vec x)\Biggr]\label{bog}
\end{equation}
where the extra cross term is a total divergence and vanishes upon integration
and:
\begin{equation}
B = {1\over 4\pi} \int d^2\vec x \; B(\vec x) 
= {1\over 8 \pi} \int d^2\vec x \;\epsilon_{ij}\vec\phi
\partial_i\vec\phi\times\partial_j\vec\phi.\label{nbbar}
\end{equation}
This expression shows that the energy of a solution is always larger or equal 
to $4 \pi\mu|B|$, giving the Bogomolnyi type bound of our model:
\begin{equation}
E = 4 \pi\mu|B|\qquad \hbox{if and only if}\qquad {1\over 2}\;
\epsilon_{ij}\,\partial_i\vec\phi\times 
\partial_j\vec\phi\pm \mu(\hat n-\vec\phi) = 0.\label{bogol}
\end{equation}

The plan of this paper is as follows. We start with an analysis of the 
Skyrmion type solution, including its
quantum rotational spectrum taking into account the deformation due to the
centripetal acceleration. To our knowledge the back reaction of the rotation on
the form of the soliton has never been taken into account in obtaining the
quantum rotational spectrum. Such an analysis would be most interesting for the
Skyrme model of nucleons. We can do so because
we can analytically solve the full equations of a steadily rotating 
baby-Skyrmion. Afterwards, we continue with our string solutions, already 
exhibited
in our previous letter\cite{7} and present closed string solutions which are 
physically
more realistic. We then present some indications on the low energy scattering
of baby-Skyrmions.
We end with a section where we consider our model on curved space.

\section{The baby-Skyrmion}

The configuration space of the model is comprised of maps from the plane $R^2$
to the target space $S^2$. Taking coordinates $\Theta$, $\Phi$ on the target
sphere (corresponding to the usual spherical polar coordinates) the best known
solution is the rotationally symmetric solution given by:
\begin{equation}
\begin{array} {l}
\Theta = f(r)
\\
\Phi = N \theta
\end{array} \label{angles}
\end{equation}
where $r,\theta$ are the usual polar coordinates on the plane. This yields the
ansatz\cite{9}
\begin{equation}
\vec\phi = (
\sin{f(r)}\cos{N\theta},\sin{f(r)}\sin{N\theta},\cos{f(r)})\label{BS}
\end{equation}
where $N$ is an integer, and is equal to the baryon number of the 
configuration. $f(r)$ has to be $\pi$ at the origin and 0 at infinity to 
obtain $B=N$.

Replacing (\ref{BS}) into the energy yields the equation of motion:
\begin{equation}
N^2 f''(r) {\sin^2{f(r)}\over r} + N^2 f'(r)^2 {\sin{2 f(r)}\over 2 r} - 
N^2 f'(r)
{\sin^2{f(r)}\over r^2} - \mu^2 r \sin{f(r)}=0\label{fBS}
\end{equation}
This complicated non-linear equation, quite surprisingly, can be integrated
analytically. By factoring $\sin{f(r)}$ we see that the first 3 terms of the
equation are a total derivative. This yields the equations (integrated)
\begin{equation}
\sin{f(r)} = 0\label{sBS1}
\end{equation}
\begin{equation}
1 - \cos{f(r)} = {\mu^2 r^4\over 8 N^2} + {a r^2\over 2} + b\label{sBS2}
\end{equation}
where $a$ and $b$ are integration constants.

Finiteness of energy and integer baryon number $B$ implies that $f(r)$ has to be
0 far from the origin. Usually this is achieved via an exponential or some
inverse power of $r$. In our case, this limit has to be attained in an
unexpected fashion. For some $r=r_0$, $f$ becomes exactly zero. This is possible
if $a\le - \sqrt{2 b}\,\mu |N|$ (reproducing in (\ref{sBS2}) the polynomial 
generally
known as representing a mexican hat). $f(0)=\pi$ imposes that $b=2$ hence the
critical value of $a$ is $-2 \mu |N|$. The function $f(r)$ is defined by
(\ref{sBS2}) only for $r\le r_0$. For $r\ge r_0$ the function 
obeys equation (\ref{sBS1}): $f$ is zero. This effectively ``patches" the 
vacuum to the
exterior of the soliton. To determine the values of $a$ and consequently $r_0$,
we minimize the energy. We note that since the soliton is of finite size, and
neither $f(r)$ nor $f'(r)$ diverge anywhere, the energy of the baby-Skyrmion is
finite. Physical intuition, verified by explicit computation, tells us that since
the energy density is a function of $(1-\cos{f})$ and its first derivative, the
total energy is minimum if $(1-\cos{f})$ attains zero at $r_0$ with a zero
derivative. This gives the values $a=-2 \mu |N|$ and $r_0=2 \sqrt{|N|/\mu}$ and
defines $f(r)$ by
\begin{equation}
f(r): \Biggl\{
\begin{array} {l}
1 - \cos{f(r)} = {\mu^2 r^4\over 8 N^2} - {\mu r^2\over |N|} + 2\qquad\hbox{
if}\quad r\le 2\sqrt{{|N|\over \mu}}
\\
0 \;\qquad\qquad\qquad\qquad\qquad\qquad\qquad\hbox{if}\quad r> 
2\sqrt{{|N|\over \mu}}
\end{array}\label{sBS}
\end{equation}
The energy of the Skyrmion can be calculated analytically:
\begin{equation}
E = {4\over 3} \;4 \pi \mu |N|.\label{EBS}
\end{equation}
This does not saturate the Bogomolnyi bound of the model. The situation is analogous to the Skyrme model in 3+1 dimensions where it is 
also possible to find a Bogomolnyi bound, which is not staurated by the 
minimum energy configuration. There, however, it is possible to define the 
model on $S^3$, which allows one to take advantage of the natural geometrical 
interpretation of the energy functional to saturate the bound by fixing the 
size of the $S^3$. No such interpretation is possible here, as the mass term 
destroys the symmetry of $S^2$ hence ruining the geometric interpretation. We 
show this in detail in the last section.

The fact that the
energy is proportional to the baryon number of the solution implies that the
energy of 2 sufficiently separated baby-Skyrmions of baryon number $B_1$ and 
$B_2$ is
equal to the energy of a localized lump of baryon number $|B_1|+|B_2|$. This 
gives support
to the possibility that there exists a sequence of deformations transforming 
the
two configurations into one another for the case ($B_1,B_2>0$). This would 
make 
possible an analytical
analysis of low energy baby-Skyrmion scattering. We will say 
more about scattering later in this paper.

We end this section on static baby-Skyrmions by computing their area. We find
trivially:
\begin{equation}
A = {4 \pi |N|\over \mu}\label{ABS}
\end{equation}
which is also proportional to the baryon number. So the area of a $|B_1|+|B_2|$
baby-Skyrmion is equal to the sum of the individual areas of the $|B_1|$ and 
$|B_2|$
baby-Skyrmions. This, together with the energy considerations given above, also
points towards a baby-Skyrmion baby-Skyrmion scattering possibly parametrized 
by a sequence of area preserving diffeomorphisms.

\section{Spinning baby-Skyrmion}

We now consider the problem of a baby-Skyrmion spinning (about the $z$ direction)
and the computation of the corresponding (semi-classical) quantum spectrum. The
usual assumption in this type of calculation is that the soliton is sufficiently
``stiff" as not to deform enough to significantly modify the energy spectrum.
This ``rigid rotor" approximation has been used frequently but seldom checked,
since this requires the ability to compute the instantaneous soliton profile 
for
any angular velocity (which is no easy task even in a purely numerical
framework). This problem was first analysed by Piette {\it et al.}~\cite{1} 
for the
full model (including kinetic term). They computed numerically the deformation
and energies of spinning baby-Skyrmions. 
We will see that our model possesses analytical solutions for
spinning baby-Skyrmions and this enables us to obtain the semi-classical
approximation to the quantum rotation energy spectrum of the baby-Skyrmion 
taking
fully into account the deformation necessary to produce the force which
maintains the centrifugal acceleration. 

The obvious ansatz for a rotating baby-Skyrmion is given by
\begin{equation}
\vec\phi = (
\sin{f(r)}\cos{(N\theta-\omega t)},
\sin{f(r)}\sin{(N\theta-\omega t)},\cos{f(r)}).\label{BSR}
\end{equation}
This ansatz has already been used by Wilczek and Zee\cite{9} to explore the 
connection between fractional spin and exotic statistics, and by Piette
{\it et al}~\cite{1}. 

We will proceed in the following manner. We first extract the instantaneous
profile of the soliton as a function of $\omega$ from the expression of the
energy which then enables us to compute (in principle) the (instantaneous)
moment of inertia and total energy of the soliton. The quantum energy spectrum
can be obtained semi-classically by using the Bohr-Sommerfeld quantization
condition which picks out the allowed values of the angular velocity $\omega$.
The $\omega$ dependence of the moment of inertia modifies the allowed
$\omega$ values (and the corresponding energy levels) from the ones obtained
with the rigid body approximation. 

Replacing the ansatz (\ref{BSR}) into the Lagrangian gives
\begin{equation}
\begin{array} {l}
L = \pi \int_0^{+\infty} dr r \sin^2{f(r)} f'(r)^2 \omega^2
\\
\qquad\qquad\qquad\qquad- \pi \int_0^{+\infty} dr \Biggl[ 
{\sin^2{f(r)}\over r} f'(r)^2 N^2 + 2\mu^2 (1-\cos{f(r)}) r\Biggr]
\end{array}\label{EBSR}
\end{equation}
after integrating over the angle $\theta$. It appears as if the only
trace of the ongoing rotation of the particle is the paramater $\omega$ in the
kinetic part of the energy, but there is of course the implicit $\omega$
dependence in $f(r)$. It is through that $\omega$ dependence of $f(r)$ that 
deformations enter the problem.

We vary $f(r)$ and obtain the following equation of motion:
\begin{equation}
\begin{array} {l}
\sin{f(r)}\cos{f(r)}f'(r)^2\Bigl({N^2\over r}-\omega^2 r\Bigr) +
\sin^2{f(r)} f''(r) \Bigl( {N^2\over r}-\omega^2 r\Bigr) -
\\
\qquad\qquad\qquad\qquad\qquad\qquad\sin^2{f(r)} f'(r) \Bigl( 
\omega^2 +{N^2\over r^2}\Bigr) -
\mu^2 r \sin{f(r)} = 0
\end{array}\label{eqBSR}
\end{equation}
As in the case of the static baby-Skyrmion, we can analytically find the
solutions of this equation:
\begin{equation}
f(r): \Biggl\{
\begin{array} {l}
1 - \cos{f(r)} = {\mu^2\over 2}\Bigl[ - {r^2\over 2 \omega^2} + 
{({r_\omega}^2\omega^2-N^2)\over 2 \omega^4} 
\ln\Bigl(1-{r^2\omega^2\over N^2}\Bigr)\Bigr] + 2
\qquad\hbox{if}\quad r\le r_\omega
\\
0
\qquad\qquad\qquad\qquad\qquad\qquad\qquad\qquad\qquad\qquad\qquad\qquad\;\;\;
\hbox{if}\quad r> r_\omega.
\end{array}\label{sBSR}
\end{equation} 
The constants of integration have been fixed by imposing $f(0)=\pi$ and
$r_\omega$ becomes the radius of the soliton and it is fixed as a function of
$\omega$, $N$ and $\mu$ by the equation
\begin{equation}
-{r_\omega^2 \mu^2\over 4 \omega^2} + {\mu^2\over 4\omega^4} 
( r_\omega^2 \omega^2 -N^2) \ln\Bigl( 1- {r_\omega^2\omega^2\over N^2}\Bigr) 
+ 2 = 
0\label{rBSR}
\end{equation}
We note that even if time dependence generates a kinetic term in the 
Lagrangian which breaks the invariance under area preserving diffeomorphisms, 
we
still obtain analytical solutions to the problem (with the exception of finding
numerically the value of $r_\omega$ for given values of $\mu$, $\omega$ and 
$N$).

When $\omega$ goes to zero we regain the static solution (\ref{sBS}). 
$r_\omega$
as a function of $\omega$ has to be obtained numerically but this does not
present any problem since $r_\omega$ is a ``well behaved" decreasing function
of $\omega$. Figure 1 shows the function $f(r)$ for several values of 
$\omega$. As $\omega$ increases $r_\omega$ decreases moderately until 
$\omega$ reaches $\omega_{max}= \sqrt{ N\mu/ 2 \sqrt{2}}$ where $r_\omega= 
r_{max} = \sqrt{ 2 \sqrt{2} N/\mu}$. After this point there does not exist a 
baby-Skyrmion type solution. $f'(r)$ is discontinuous at $r_{max}$.

We compute the classical energy $E$ as a function of $\omega$ and compare it 
with
the same quantity in the case of the rigid body. The dynamical moment of 
inertia
is defined by
\begin{equation}
I(\omega) = 2 \pi \int_0^{+\infty} dr\; r \sin^2{f(r)} f'(r)^2\label{MI}
\end{equation}
which is $2\; T/\omega^2$ where $T$ is the kinetic energy of the system (see 
the
first term in (\ref{EBSR})). Replacing for $\sin{f(r)} f'(r)$ with the 
solution 
(\ref{sBSR}) and making the change of variables
\begin{equation}
\begin{array} {l}
r = \sqrt{{N\over \mu}} \rho
\\
\omega = \sqrt{N \mu} w
\end{array}\label{chvar}
\end{equation}
we obtain
\begin{equation}
I(\omega(w)) = {\pi\over 2}\int_0^{\rho_w} d\rho \;\rho^3\; \Biggl( 
{\rho^2-\rho_w^2\over 1 - w^2 \rho^2} \Biggr)\label{Iwrho}
\end{equation}
where $\rho_w$ is a solution of the equation (obtained from (\ref{rBSR}))
\begin{equation}
\rho_w^2 w^2 + (1-\rho_w^2 w^2) \ln {(1-\rho_w^2 w^2}) - 8 w^4 = 0\label{cwrho}
\end{equation}
and which is independent of $N$ and $\mu$. $\rho_w$ ranges from 2 to 
$\sqrt{2 \sqrt{2}}$ while $w$ ranges from 0 to $w_{max}=1/\sqrt{2 \sqrt{2}}$.
The moment of inertia of the static baby-Skyrmion 
is defined as 
\begin{equation}
I_0 = \lim_{\omega\rightarrow 0}{I(\omega)}.\label{I0}
\end{equation}
This amounts to replacing the static solution for $f(r)$ in (\ref{MI}) and the
result $I_0=16\pi/3$. $I_{max} = I(\omega_{max}) = 8\pi$. The 
energy of a rigidly rotating baby-Skyrmion is given by
\begin{equation}
E_N(\omega) = {4\over 3} 4 \pi \mu |N| + {8 \pi\over 3} \omega^2\label{ERR}
\end{equation}
while the true rotating baby-Skrymion's energy is:
\begin{equation}
\begin{array} {l}
E_N(\omega) = {\pi \mu^2\over 16 \omega^2} \Biggl[ 112 N^2 - 7\mu^2 
r_\omega^4 - 16 r_\omega^4 \omega^2\Biggr]
\\
\qquad\quad= {\pi \mu |N| \over 16 w^2} \biggl[ 112 - 7 \rho_w^4 - 16 
\rho_w^2 w^2\biggr]
\end{array}\label{EBSRr}
\end{equation}
which is linear in $\mu$ and $|N|$.
Figure 2 compares the rotational energy of a rigid baby-Skyrmion and the energy
of the deformable baby-Skyrmion, and we note that the true energy is larger 
than the rigid rotor approximation
for all values of $\omega$ in accord with the behaviour found
by Piette {\it et al}~\cite{1}. This behaviour is as expected since, for 
example, a block of 
rubber would react to a steady rotation by increasing its energy faster than 
a rigid rotor. The surprising behaviour that we find is that $r_\omega$ is a 
decreasing function of $\omega$ and there is a $\omega_{max}$ after which 
there is no solution of this type. Even though the size of the baby Skyrmion 
decreases, its energy and moment of inertia increase. Their densities tend to 
concentrate towards the outside (see Figure 3). We speculate that this is a 
precursor to 
breakup or the emission of some form of radiation. The energy density actually becomes discontinuous at $r_{max}$ for $\omega = \omega_{max}$.
 
The general form of the energy
of a spinning deformable body is written as
\begin{equation}
E(\omega) = {1\over 2} I(\omega)\, \omega^2 + U(\omega) + M\label{CD}
\end{equation}
where $U(\omega)= V - M$, represents the potential energy stored in the 
particle as it reacts to centrifugal forces, $M = V|_{\omega\rightarrow0}$ is
the mass of the static baby-Skyrmion (\ref{EBS}) and
\begin{equation}
V = \pi \int dr \Biggl[ \sin^2{f(r)} f'(r)^2 {N^2\over r} +
2 \mu^2 r \Bigl(1 - \cos{f(r)}\Bigr)\Biggr].\label{VBSR}
\end{equation} 
In the case of the
rotating physical baby-Skyrmion, this is a slowly increasing function of
$\omega$. 

To compute the quantum rotational spectrum of the baby-Skyrmion we use the
Bohr-Sommerfeld condition\cite{10} on the action and energy of the system:
\begin{equation}
\begin{array} {l}
W(E_n) = S[ \tau(E_n)] + E_n \tau(E_n)
\\
\qquad\qquad= (2 n + 1)\pi\hbar
\end{array}\label{BhS}
\end{equation}
where $\tau(E_n)$ is the period of the motion for a given energy $E_n$ and $S$
the corresponding action. Due to the stationary nature of the soliton, only the
kinetic term contributes in (\ref{BhS}), yielding the quantisation condition
\begin{equation}
\omega_n = {1\over 2 I(\omega)} (2 n + 1) \hbar\label{QC}
\end{equation}
$\omega_n$ is then obtained numerically as a root of equation (\ref{QC}) and 
then
replaced in the expression of the energy. It is in the solution of equation
(\ref{QC}) that the deformation of the soliton ``interacts" with its rotation,
representing the action of the system which chooses its shape to achieve a
quantum state. 

In the case of the rigid body, the energy rises as $n^2$. 
In taking account of deformations,
$I(\omega)$ is larger than $I_0=16\pi/3$, making the $\omega_n$ smaller than 
the corresponding rigid body solution. Then the net effect of the deformation 
will be
to shorten the gap between sequential energy levels. We show both spectra
in Figure 4. We observe that the rigid body approximation is acurate for the 
first few levels, followed by a marked divergence. Since there does not exist 
baby-Skyrmion type solutions for $\omega\ge \omega_{max}$ the rotational 
energy spectrum of Figure 4 contains a finite number of quantum levels for the 
soliton. The number of those levels depends on $\mu$ and $N$ and is easily 
shown to be:
\begin{equation}
n_{max} = \hbox{\rm Integer Part}\Biggl[\,{1\over 2}\,
\biggl(16\,\pi\,\sqrt{{N\mu\over 2\sqrt{2}}} - 1\biggr)\Biggr).
\label{eqmmax} 
\end{equation}
In the case of the full baby-Skyrme model, Piette {\it et al.}~\cite{1} found 
the
classical rotational spectrum of the baby-Skyrmion. They showed that two 
regimes
existed according to the relative size of the angular velocity $\omega$ and 
pion
mass $\mu$: if the baby-Skyrmion spins too quickly the solution of the equation
of $f(r)$ becomes oscillatory at infinity, rendering the baby-Skyrmion's energy
and moment of inertia infinite. They also showed that a baby-Skyrmion
``cranked" that high will quickly reduce its energy and angular velocity by
emitting pion radiation until $\omega\stackrel{<}{\sim}\mu$, which is the 
``steady rotation" regime. In our case, the size of the baby-Skyrmion 
actually decreases with $\omega$ but
the polynomial profile prevents any oscillatory or long range behaviour of
the function $f(r)$. There is an $\omega_{max}$ after which there is no 
solution of this form.

We end this section by stressing that the quantum states so obtained are not
eigenstates of spin $\vec S$, nor isospin $\vec I$, but of the sum $\vec S + 
\vec
I$ of these operators. This is due to the fact that the
baby-Skyrmion is only invariant under the diagonal group spin/isospin. Thus 
the tower of states constructed would be exactly the same if we had
considered iso-rotations about the 3 axis. The quantization we have followed
treats the solitons as bosons. A more general quantization as anyons is
possible, as put forward by Wilczek and Zee\cite{9} and requires the inclusion 
of a Hopf term. We will not elaborate on this possibility here.

\section{Closed string in 2+1 dimensions}

In a previous article\cite{7} we introduced the ansatz for extended solutions 
resembling
strings or strips, in the complete baby-Skyrme model and related models. Here 
we
shall construct a physical state with such strings and compute its energy. 
The simplest ``physical state" one can construct with the string is an annulus
made by bending the string and closing it. We refer the reader to ref. [7] 
for the details and notations. To do this, one has to re-interpret
the coordinate along the string, $y$, as an angle which winds around the 
origin. The $y$ axis is
mapped to a circle of radius $A$ and the $x$ coordinate becomes radial. The
ansatz that we use is
\begin{equation}
\vec\phi_< = (\sin{f_<(r)} \cos{N\theta},
- \sin{f_<(r)} \sin{N\theta},
\cos{f_<(r)})\quad\hbox{for}\quad r<A\label{AP}
\end{equation}
\begin{equation}
\vec\phi_> = (\sin{f_>(r)} \cos{M\theta},
\sin{f_>(r)} \sin{M\theta},
\cos{f_>(r)})\quad\hbox{for}\quad r>A\label{AG}
\end{equation}
where $\vec\phi_<$ represents the inner part of the annulus ($r<A$) and
$\vec\phi_>$ the outer part ($r>A$). We have deliberately generalised the 
ansatz by
winding $\vec\phi_<$ $N$ times and $\vec\phi_>$ $M$ times around the origin.
There is no boundary problem since $f(r=A)=\pi$ no matter the values of $N$ and
$M$. The baryon number of the annulus is the sum $N+M$ where $N$ and $M$ are
both non-zero integers (although their sum can be zero). 

We now compute the energy and profile for the annulus. The procedure to follow
is exactly the same as before with the exception that we have two functions to
compute: one, $f_<$, for the inner part of the annulus, and the other, $f_>$,
for the outer part. The corresponding differential equations are coupled
together by the parameter $A$. One finds the ``inner" solution $f_<$:
\begin{equation}
f_<(r): \Biggl\{
\begin{array} {l}
1 - \cos{f_<(r)} = {r^4 \mu^2\over 8 N^2} + r^2 {\mu^2\over 4 N^2}\Bigl({4
|N|\over \mu} - A^2\Bigr) + {A^4 \mu^2\over 8 N^2} - {\mu A^2\over |N|} + 2
\;\hbox{if}\; A\ge r\ge r_< 
\\
0 \qquad\qquad\qquad\qquad\qquad\qquad\qquad\qquad\qquad\qquad\qquad\quad\;
\,\quad\hbox{if}\;\;\; r < r_<
\end{array}\label{sAP}
\end{equation}
where $r_< \equiv \sqrt{A^2 -{4 |N|\over \mu}}$ and the ``outer" solution $f_>$:
\begin{equation}
f_>(r): \Biggl\{
\begin{array} {l}
1 - \cos{f_>(r)} = {r^4 \mu^2\over 8 M^2} - r^2 {\mu^2\over 4 M^2}\Bigl({4
|M|\over \mu} + A^2\Bigr) + {A^4 \mu^2\over 8 M^2} + {\mu A^2\over |M|} + 2
\;\hbox{if}\; A\le r\le r_> 
\\
0 \qquad\qquad\qquad\qquad\qquad\qquad\qquad\qquad\qquad\qquad\qquad\quad\!
\!\qquad\hbox{if}\;\;\; r > r_>
\end{array}\label{sAG}
\end{equation}
where $r_> \equiv \sqrt{A^2 +{4 |M|\over \mu}}$. This solution has energy
\begin{equation}
E_{N,M} = {4\over 3} 4 \pi \mu \Bigl(|N|+|M|\Bigr)\label{EANM}
\end{equation}
and area
\begin{equation}
\begin{array} {l}
A_{N,M} = \pi \,(r_>)^2 - \pi \,(r_<)^2 
\\
\;\;\,\qquad= {4 \pi\over \mu}\Bigl(|N|+|M|\Bigr)
\end{array}\label{AANM}
\end{equation}
which is independent of the parameter $A$, for the same reason that the area of
the string was independent of its length $L$. Indeed a large radius $A$ implies
a very thin annulus while a smaller $A$ implies a thicker annulus. Finally
there is a minimum $A$ for which $r_< = 0$ and the annulus becomes a disc
however with a different configuration inside with respect to the 
baby-Skyrmion.

Before ending this section let us
make a few remarks. First, the energy of the solution does not change if we
``move" the baryon number from the inner part of the annulus to the outer as 
long as the total $N+M$ does not change (for both $N$ and $M$ positive). This 
generates another degeneracy in the energy spectrum. 

Second, in order for $r_<$ to be real, one has to ensure that $A^2\ge 4 |N|/\mu$,
implying that a solution of large baryon number must have a large radius $A$, or
that most of its baryon number (and energy) must be stored in the outer shell
of the annulus.

Third, the state with $M+N$ is different from the baby-Skyrmion with $B=N+M$ in
that it actually contains the vacuum in its interior: $f_<=0$ for $0<r<r_<$. We
see that in the $B=2$ sector for example, we have 2 different lumps degenerate
in energy: the $N=2$ baby-Skyrmion, or ``deuteron", and the $N=M=1$ annulus. We
remind the reader that the deuteron cannot be considered analogous to the 
annulus
since $f=\pi$ at its center. We will give an indication in the next section as
to the importance of these states in baby-Skyrmion scattering.

\section{Considerations on soliton scattering}
 
We use the method of Manton\cite{11} while exploring the symmetries of the 
deuteron
state in the usual 3+1 dimensional Skyrme model and the symmetries of scattering
states. His method was based on the fact that
looked from a distance, a Skyrmion resembles a triad of orthogonal dipoles.
Piette {\it et al}~\cite{1,3}
showed that similarly baby-Skyrmions look like a pair of
orthogonal dipoles. One way to see this is to 
use the $\vec\phi=(\phi^1,\phi^2,\phi^3)$
parametrisation and the constraint $\vec\phi^2 =1$. Using the constraint to
eliminate, say, $\phi^3$ one can describe a baby-Skyrmion
configuration completely
with the fields $\phi^1$ and $\phi^2$. Let us now consider the
region of the plane defined by $x\sim 0, y>0$. Since the baby-Skyrmion field is
radial when projected in the $(1,2)$ plan, in this region $\phi^1$ will be
very small, while $\phi^2$ will dominate and be positive. In the opposite region
($x\sim 0, y<0$), it is the same situation but now $\phi^2$ dominates while
being negative. So the field $\phi^2$, which falls at infinity as an
exponential in the full model, represents from a far, a dipole oriented
parallel to the $y$ axis. The same is valid for $\phi^1$ except along the $x$ 
axis. The
baby-Skyrmion is thus constructed from a pair of orthogonal dipoles (see 
\cite{2,3}
for a more rigorous demonstration of the dipole nature of the field of the
baby-Skyrmion). 

It was Manton's\cite{11} idea to show that the natural evolution of these 
dipoles 
during
a Skyrmion-Skyrmion scattering with relative orientation 180$^\circ$ 
(the so-called
most attractive channel) is compatible with a 90$^\circ$ scattering. This 
was based on
the observation that the discrete symmetries under reflection of the Skyrmion
fields were conserved throughout the scattering. He also put forward the
hypothesis that the deuteron state, which is the expected intermediate state of
the scattering, possesses a cylindrical symmetry around the axis perpendicular
to the scattering plane, englobing the discrete symmetries mentioned. 

A similar reasoning should also be valid in the full baby-Skyrme model, and
numerical simulations have concurred with the 90$^\circ$ scattering and an 
intermediate state made up of a $N=2$ baby-Skyrmion for the case of 2
baby-Skyrmions with a relative iso-rotation of $180^\circ$ around the 
$z$ axis\cite{1}.
This is also completely consistent with Mackenzie's\cite{12} theorem which 
proves 
under very general and reasonable assumptions that head on, geodetic scattering
of identical particles must be through 90$^\circ$ if the coincident point is
attained. 
Scattering with different relative orientations has also been studied, as we
will discuss in more detail later.

In our case, we do not have any dipole picture since our solitons have finite
size and the ``long range" interactions are not well approximated by a linear
theory of massive or massless pions. What we have here are well determined
regions where $\phi^1$ or $\phi^2$ dominate clearly (we represent these regions 
by a circle containing a ``+" or ``-" sign, as in Manton's notation. See Figure
5 for a picture of a baby-Skyrmion, and 
one that has been rotated by $180^\circ$
around the $z$ axis). This does
not keep us from using Manton's framework. If anything, the finite and exactly 
free
nature of our solitons render exact for a pair of baby-Skyrmions the analog of 
the approximate, discrete symmetries of a
pair of usual Skyrmions, even when the baby-Skyrmions are at a finite distance 
from one 
another (in fact, anytime that they don't actually touch each other).

We now turn to the scattering of baby-Skyrmions without a relative 
iso-rotation.
This case was not discussed by Manton, but treated numerically by Piette 
{\it et al.}~\cite{1} for the full baby-Skyrme model. They find 0$^\circ$ or 
180$^\circ$ scattering. These cases are actually indistinguishable from just 
the
scattering data! Figure 6 shows the ``dipole image" of a pair of 
baby-Skyrmions without
any relative iso-rotation for both fields $\phi^1$ and $\phi^2$. One can easily
verify that this configuration transforms according to 
\begin{equation}
\begin{array} {l}
x\rightarrow -x: \phi^1 \rightarrow -\phi^1 \qquad\qquad y \rightarrow -y: \phi^1
\rightarrow \phi^1
\\
\qquad\qquad\;\phi^2 \rightarrow \phi^2\qquad\qquad\qquad\qquad\;\;\;\,\phi^2
\rightarrow -\phi^2
\end{array}\label{SymLAC}
\end{equation}
which are actually just the symmetries of a single baby-Skyrmion. These 
symmetries are not
shared by a pair of baby-Skyrmions moving away along the $y$ axis so we conclude
that the most probable scattering process is 180$^\circ$/0$^\circ$ 
collisions. But
what could be the intermediate state of the collision? We know that it cannot be
the deuteron since it does not transform according to (\ref{SymLAC}). In fact
the only $B=2$ state we know of which possesses the right symmetries is the
$B=2$ annulus described in the preceding section, wich is degenerate in energy
with both the deuteron and a pair of separated $B=1$ baby-Skyrmions. We might
then speculate that this annulus could be the intermediate state arising in
the 180$^\circ$ or 0$^\circ$ scattering. It would be interesting to 
actually compute the
energy of this annulus state in the full baby-Skyrmion model and to look for
this state by studying numerically the scattering between unrotated
baby-Skyrmions.

\section{Baby-Skyrme type model on a curved space}

We end this article by briefly studying the properties of the baby-Skyrmion of
our model when put on a curved space, namely on a 2-sphere of radius $R$
parametrized by the angles $\theta$ and $\phi$ where $\theta\in[0,\pi]$ and
$\phi\in[0,2\pi]$. Our aim is to study the properties of a dense system of 
baby-Skyrmions, like when put on a regular 2-dimensional array. The same
procedure was followed by N.S. Manton {\it et al}~~\cite{13} in the case of 
the usual Skyrme model in 3+1 dimensions without a mass term, where ordinary
space was compactified into a 3-sphere. They showed that below a critical
radius of the 3-sphere, the solution of lowest energy is not the Skyrmion
but a configuration with uniform energy density, and recognized there chiral
symmetry restoration. This makes sense since the Skyrme model is related 
to QCD in the limit of a large number of colours. 

We will now compute the energy of the baby-Skyrmion for decreasing values of
the 2-sphere radius, and compare it to the energy of the configuration with
constant energy density. 

When replacing the ansatz for the baby-Skyrmion on a 2-sphere
\begin{equation}
\vec \Phi(\theta,\phi) = \biggl(\sin f(\theta) \cos\phi, \sin f(\theta) \sin
\phi, \cos f(\theta)\biggr)\label{sks2}
\end{equation}
into the equation of motion of the Lagrangian on a space-time of
metric $g_{\mu\nu}$ (in our case $\sqrt{g}=R^2 \sin\theta$) given by
\begin{equation}
\partial_\mu\Biggl( \sqrt{g} \partial^\nu \vec\Phi \vec\Phi\cdot
\biggl(\partial^\mu \vec \Phi\times\partial_\nu \vec\Phi\biggr)\Biggr)
+ \sqrt{g} \mu^2 \vec\Phi\times\hat n =0\label{eqg}
\end{equation}
we find the following analytical solution:
\begin{equation}
f(\theta): \Biggl\{
\begin{array} {l}
1 - \cos{f(\theta)} = {R^4 \mu^2\over 2 N^2} ( \cos{\theta_0}-\cos{\theta})^2
\qquad\hbox{if}\quad \theta\le \theta_0
\\
0 \;\qquad\qquad\quad\qquad\qquad\qquad\qquad\qquad\qquad\!\hbox{if}\quad \theta>\theta_0
\end{array}\label{bss2}
\end{equation}
where $\theta_0$ is the ``size" of the solution on the 2-sphere and is given by
\begin{equation}
\sin{\theta_0\over 2} = {1\over R} \sqrt{|N|\over \mu}.\label{tailles2}
\end{equation}
This solution has again compact support and finite size and can only exist for 
$R \sqrt{\mu/|N|} \ge 1$ in order for $\theta_0$ to be
real. $R \sqrt{\mu/|N|} = 1$ represents the situation where the soliton 
entirely covers the
2-sphere and continues to satisfy $\sin{f(\theta_0)} f'(\theta_0)$ equal to 0 
with $\theta_0=\pi$.
The energy of the solution is obtained by replacing equation (\ref{bss2}) in 
the general expression for the energy in curved space. We find: 
\begin{equation}
E = {4\over 3} 4\pi \mu|N|\label{ebss2}
\end{equation}
which is surprisingly independent of $R$, and hence, equal to the
energy of the baby-Skyrmion in flat space. 

If we want to reduce still the radius of the 2-sphere, we have to relax the
constraint that $\sin{f(\theta)} f'(\theta)$ be 0 at the edge of the soliton.  
The derivative will then be discontinuous at that point but the
energy, being only function of the first derivative of $f(\theta)$, will not
diverge. The so-called ``compressed" baby-Skyrmion solution, existing only for
$R \sqrt{\mu/|N|} \le 1$, is also obtainable
analytically:
\begin{equation}
1 - \cos{f(\theta)} = - {R^4 \mu^2\over 2 N^2} \sin^2{\theta} + 
2 \cos^2{\theta\over 2}
\qquad\quad \theta\in [0,\pi]\label{bscs2}
\end{equation}
This solution completely fills the 2-sphere and has energy:
\begin{equation}
E = {2\pi N^2\over R^2} + 4\pi \mu^2 R^2 - {2\pi \mu^4 R^6\over 3}\label{ebscs2}
\end{equation}
which diverges if $R\rightarrow 0$.  The former
happens because of the 4 derivatives of the Skyrme term combined with the
integration measure and illustrates that a point soliton will have infinite
energy.  Of course, the energy of the two solutions will become equal in the 
limit where $R \sqrt{\mu/|N|}=1$.

We now want to compare the energy of these solutions with the energy of the
configuration with uniform energy density. Because of the presence of the mass
term in the Lagrangian, this type of configuration is not a solution of the
equations of motion. This was not the case in the work of Manton 
{\it et al}~\cite{13}
since they did not include any mass term to stabilize the solitons
against scale changes of the coordinates.
In our case, with $f(\theta)= \pi - \theta$ and the baby-Skyrmion ansatz, we
find the following energy:
\begin{equation}
E = {2\pi N^2\over R^2} + 4\pi \mu^2 R^2\label{eisos2}
\end{equation}
which is always higher in energy than the baby-Skyrmion. We see that there
is no phase transition between the two states as $R$ changes values. 
 
\acknowledgments
We thank R. Mackenzie and W.J. Zakrzewski for useful discussions, W.C.
Chen for help with the numerical work, and the referee for his important 
criticisms. This
work supported in part by NSERC of Canada and FCAR of Qu\'ebec.

\begin{figure}                                                               
\vspace{-2.cm}
\leavevmode
\epsfxsize=0pt\epsfbox{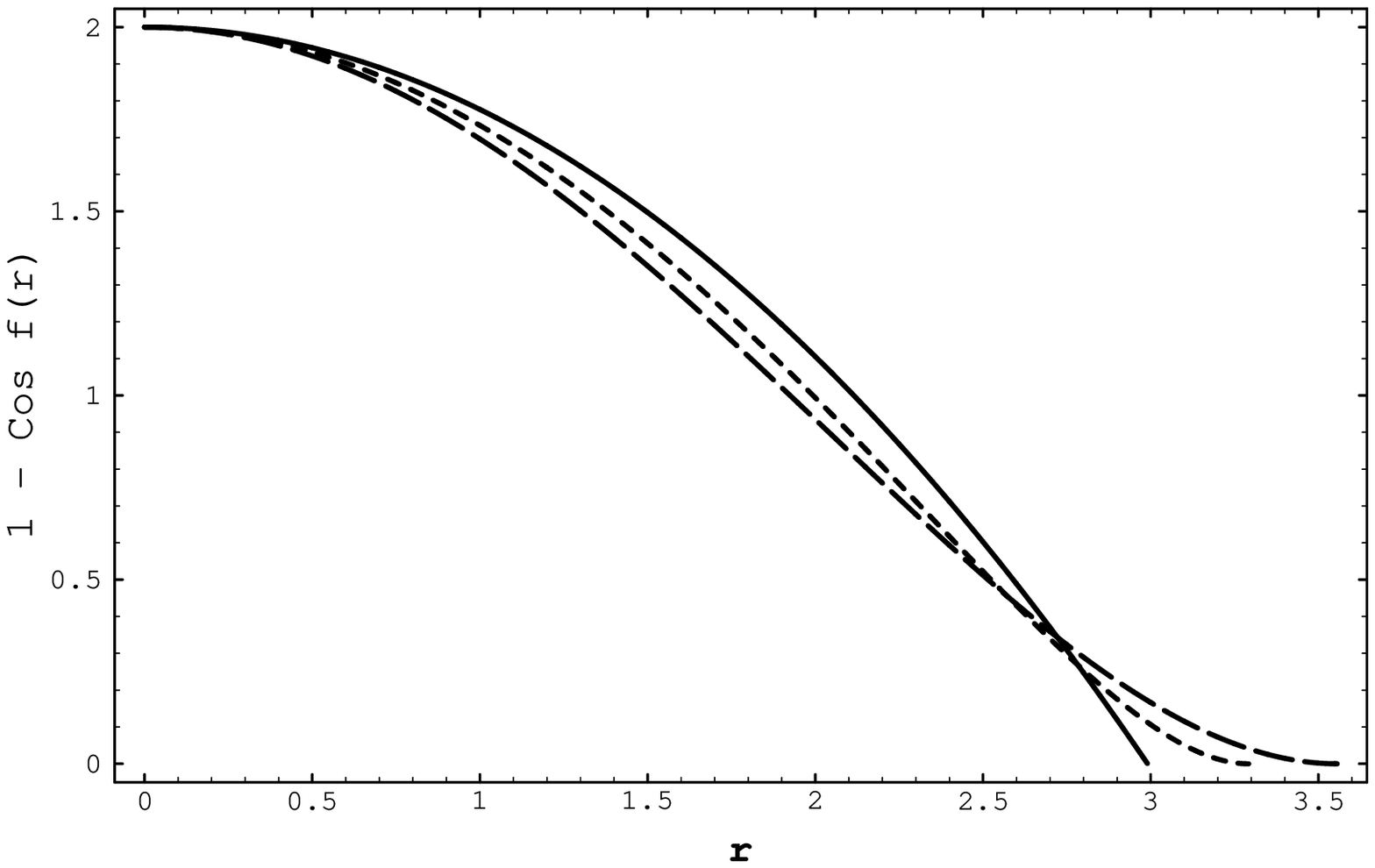}
\vspace{-2.cm}
\caption{Profile of the function $f(r)$ of a baby-Skyrmion spinning at
the angular velocity 0 (long dashed line), 0.25 (dashed line) and 
$\omega_{max}$ (solid line), with $N=1$ and $\mu^2=0.1$.}
\end{figure}
\begin{figure}
\vspace{-2.cm}
\leavevmode
\epsfxsize=450pt\epsfbox{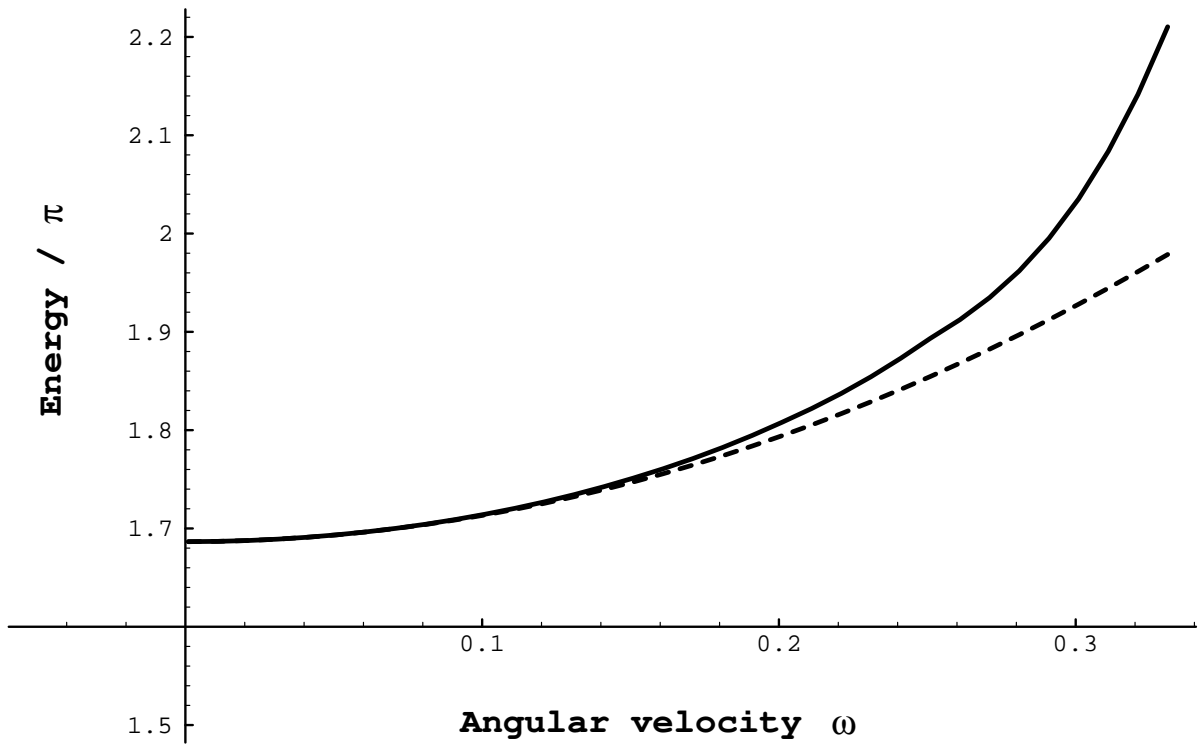}
\vspace{-2.cm}
\caption{Plots for $N=1$ and $\mu^2=0.1$ of the energies of a baby-Skyrmion
spinning at angular velocity $\omega$ ranging from 0.001 to $\omega_{max}$, 
in the rigid body
approximation (dashed line), and when centripetal deformations are taken into 
account (solid line).}
\end{figure}
\begin{figure}
\vspace{-2.cm}
\leavevmode
\epsfxsize=450pt\epsfbox{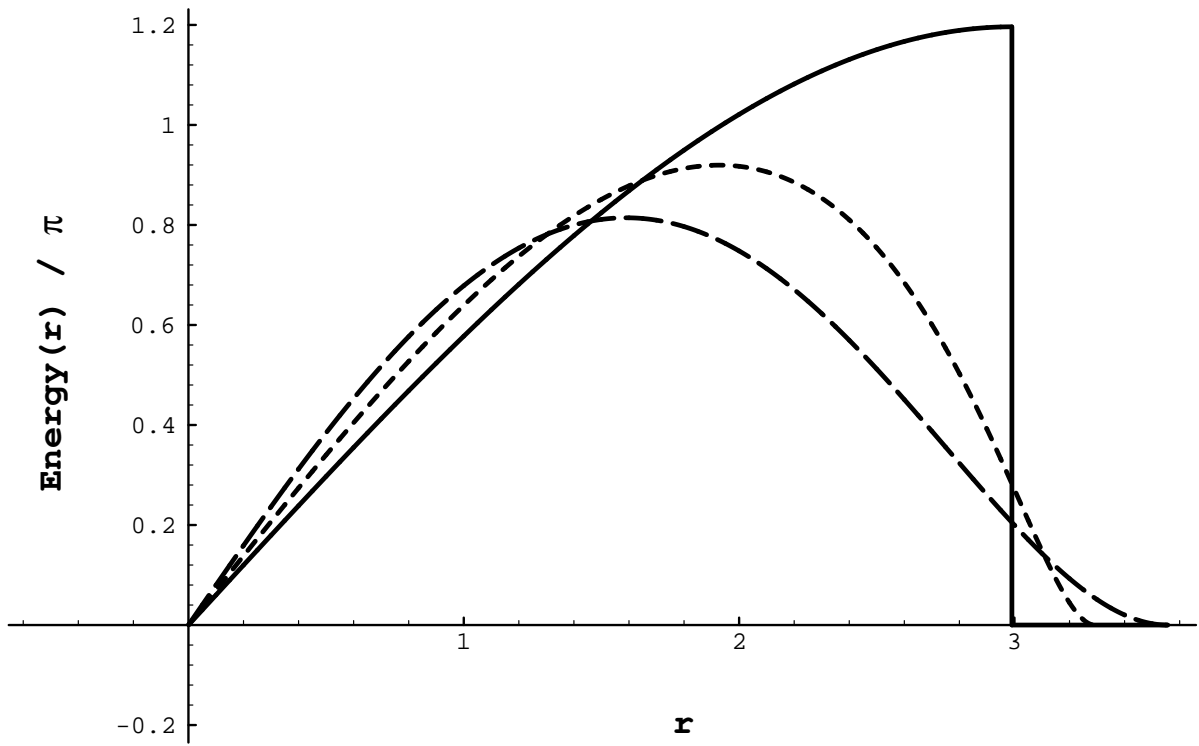}
\vspace{-2.cm}
\caption{Plot of the radial energy distribution of a baby-Skyrmion spinning at
angular velocities 0 (long dashed line), 0.25 (dashed line) and 
$\omega_{max}$ (solid line), for $N=1$ and $\mu^2=0.1$.}
\end{figure}
\begin{figure}
\vspace{-2.cm}
\leavevmode
\epsfxsize=450pt\epsfbox{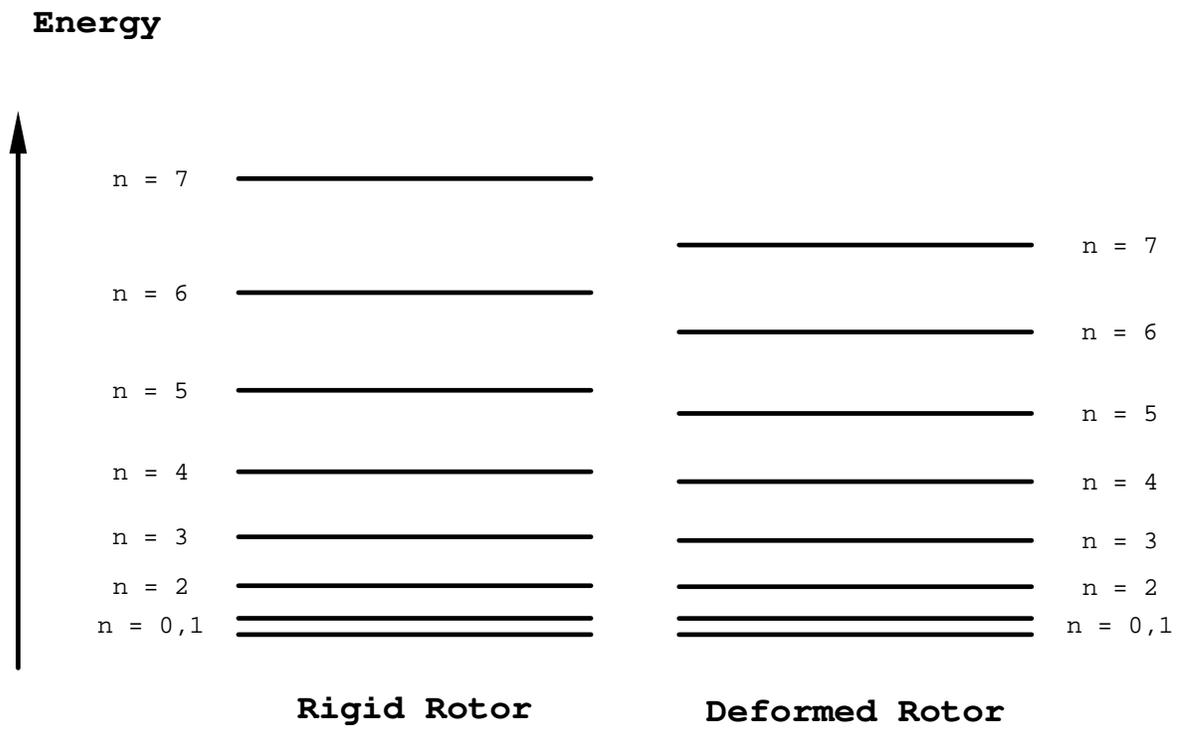}
\vspace{-2.cm}
\caption{Comparison of the quantum rotational energy spectra for a 
baby-Skyrmion 
with $N=1$ and $\mu^2 = 0.1$, in the rigid rotor approximation and for the 
exact, deformed rotor.}
\end{figure}
\begin{figure}
\vspace{-2.cm}
\leavevmode
\epsfxsize=300pt\epsfbox{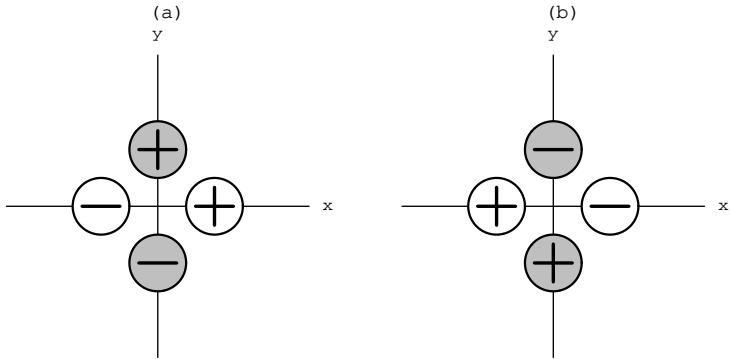}
\vspace{-2.cm}
\caption{Representation of the baby-Skyrmion field using Manton's notation of
dipole regions. The regions where the field $\phi^1$ dominates are drawn in 
white, while those where $\phi^2$ dominates are drawn in gray. The figure (a)
represents a baby-Skyrmion unrotated, while the figure (b) represents
one that has been rotated by $180^\circ$ around the $z$ axis.}
\end{figure}
\begin{figure}
\vspace{-2.cm}
\leavevmode
\epsfxsize=300pt\epsfbox{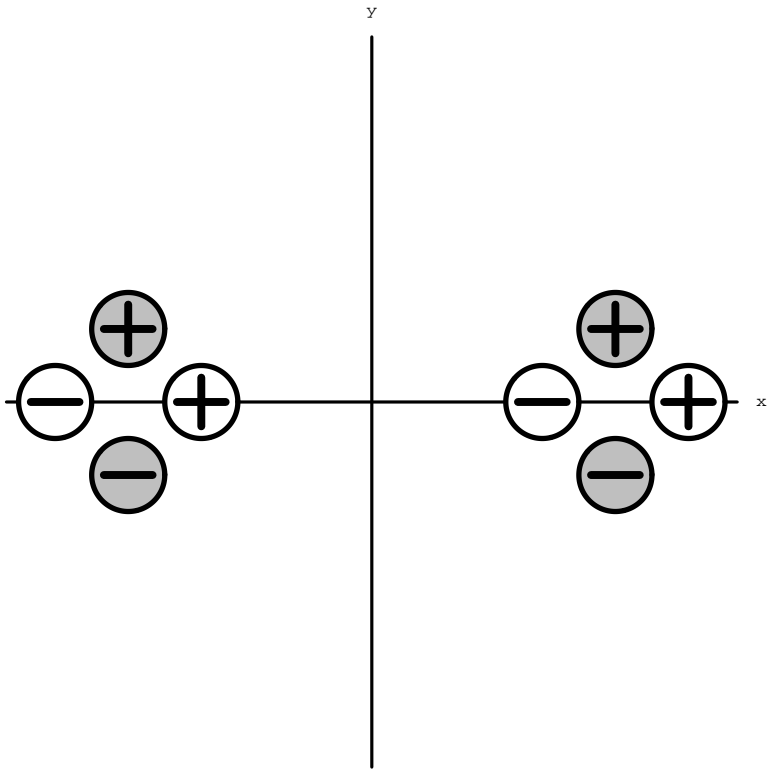}
\vspace{-2.cm}
\caption{Field of baby-Skyrmions with no relative iso-rotation.}
\end{figure}
\end{document}